\documentclass[12pt]{article}%
\usepackage{amsmath}
\usepackage{amsfonts}
\usepackage{amssymb}
\usepackage{graphicx}%
\providecommand{\U}[1]{\protect\rule{.1in}{.1in}}

\csname @addtoreset\endcsname{equation}{section}
\begin{document}
\begin{titlepage}
\begin{flushright}
\hfill DMUS--MP--15/01 \\
\end{flushright}
\vspace{.3cm}
\begin{center}
\renewcommand{\thefootnote}{\fnsymbol{footnote}}
{\Large{\bf On Non-extremal Instantons and Black Holes}}
\vskip1cm
\vskip 1.3cm
Jan B. Gutowski$^1$  and Wafic A. Sabra$^2$
\vskip 1cm
{\small{\it
$^1$Department of Mathematics, University of Surrey\\
Guildford, GU2 5XH, United Kingdom\\}}
\vskip .6cm {\small{\it
$^2$ Centre for Advanced Mathematical Sciences and Physics Department\\
American University of Beirut\\ Lebanon  \\}}
\end{center}
\bigskip
\begin{center}
{\bf Abstract}
\end{center}
We consider a general analysis and a specific ansatz for the study of non-supersymmetric
solutions in arbitrary dimensions and various metric signatures.
In all cases, we find that the conditions on the solutions
can be written in terms of quadratic forms involving the
gauge coupling of the theory and constants of integration
associated with the scalar fields.
Depending on the signature of the metric, our analysis should provide a
general framework for finding non-extremal black holes, instantons, branes and S-branes.
\end{titlepage}

\section{Introduction}

Gravitational supergravity solutions admitting fractions of supersymmetry in
various spacetime dimensions and particularly in four and five dimensions have
been a subject of intense research activities in recent years (see for example
\cite{recentlower}). The earliest systematic work in this direction is that of
Tod \cite{Tod} where solutions of Einstein-Maxwell theory, ($D=4,$ $N=2$
minimal supergravity) admitting parallel spinors were classified. The
solutions with time-like Killing vectors are the IWP solutions \cite{IWP}, for
which the static limit is the MP solution \cite{mp}. Their generalisations to
supergravity theories coupled to vector multiplets were performed in
\cite{sab, BLS}. As a consequence of \cite{ortin}, it is well-understood how
such solutions appear in the context of the general classification of
supersymmetric solutions in four dimensions. For $N=2,d=5$ supergravity with
vector multiplets, general electrically and magnetically charged string
solutions were initially considered in \cite{sabrafive, sabrachams5, mag}.
Later, a systematic classification of five-dimensional supersymmetric
solutions was considered in \cite{unique5, 5dsys}. Non supersymmetric
solutions in five dimensions were first considered in \cite{BCS} where an
explicit electrically charged solution for the so-called STU model was given.
More analysis of non-supersymmetric solutions was later presented in
\cite{nonextremaljan} which also included the study of non-supersymmetric
magnetically charged string solutions. Moreover, specific examples of
non-extremal solutions were presented in \cite{GO4}.

More recently, some attention was given to the study of supersymmetric
gravitational instanton solutions to the Euclidean Einstein equations of
motion. Recent work on gravitational instantons considered the Euclidean
analogues of the IWP metrics \cite{DH07, instantons} as well as solutions for
theories with cosmological constant. Euclidean versions of $N=2,$ $d=4$
supergravity theories have been constructed in \cite{mohaupt3} via the
dimensional reduction of $N=2,$ $d=5$ supergravity theories on a time-like
circle. Gravitational instanton solutions for these Euclidean theories were
obtained in \cite{instanton} via the analysis of the Euclidean Killing spinor
equations \cite{eusy}. These supersymmetric gravitational instanton solutions
can be thought of as the Euclidean analogues of the black hole solutions found
in \cite{BLS}.

In our present work we consider a general analysis and a specific ansatz for
non-supersymmetric solutions in arbitrary dimensions. Depending on the
signature of the metric, our present analysis should provide a general
framework for finding non-extremal black holes, instantons, branes and
S-branes. The examples given are electric and magnetic solutions in four and
five dimensions. For all these examples, all equations of motion are reduced
to two algebraic conditions which depend on the gauge coupling structure of
the theory. For electrically charged solutions, the equations are reduced to
\begin{equation}
Q^{IJ}S_{IJ}=0,\text{ \ \ \ \ }\partial_{i}Q^{IJ}S_{IJ}=0
\end{equation}
while for magnetic solutions one obtains%

\begin{equation}
Q_{IJ}\phi^{IJ}=0,\text{ \ \ \ \ \ \ }\partial_{i}Q_{IJ}\phi^{IJ}=0
\end{equation}
where $Q_{IJ}$ is the gauge coupling, $S_{IJ}$ and $\phi^{IJ}$ are related to
the constants and charges appearing in the solutions, and $\partial_{i}$
denotes variation with respect to the scalar fields of the theory in question.

We organise our work as follows. The next section contains our general
analysis for the non-extremal electric and magnetic solutions in arbitrary
dimensions. Section 3 contains examples of four-dimensional solutions for
black holes, instantons and time-dependent solutions. In section 4,
five-dimensional solutions are considered.

\section{General Solutions: Analysis of Einstein Equations}

Consider a general $d$-dimensional gravity coupled to scalar fields $\phi^{i}%
$, and $\ell$-form gauge field strengths $F^{I}$, with action
\begin{equation}
\mathcal{L}_{d}=\sqrt{|g|}\bigg(R-{\mathcal{G}}_{ij}\nabla_{M}\phi^{i}
\nabla^{M}\phi^{j}-{\frac{1}{2\ell!}}{\tilde{Q}}_{IJ}(\mathcal{F}^{I}
)_{M_{1}\dots M_{\ell}}(\mathcal{F}^{J})^{M_{1}\dots M_{\ell}}
\bigg)+\mathcal{L}_{CS} \label{act}%
\end{equation}
$\mathcal{G}_{ij}$, ${\tilde{Q}}_{IJ}$ are functions of the scalar fields
$\phi^{i}$, and $\mathcal{L}_{CS}$ is a Chern-Simons term, whose structure
depends on the theory in question.

The Einstein field equations are
\begin{align}
R_{MN}  &  =\mathcal{G}_{ij}\nabla_{(M}\phi^{i}\nabla_{N)}\phi^{j}+{\frac
{1}{2(\ell-1)!}}{\tilde{Q}}_{IJ}(\mathcal{F}^{I})_{(M|N_{1}\dots N_{\ell-1}
|}(\mathcal{F}^{J})_{N)}{}^{N_{1}\dots N_{\ell-1}}\nonumber\\
&  +{\frac{1}{2\ell!}}\bigg({\frac{\ell-1}{2-d}}\bigg){\tilde{Q}}
_{IJ}(\mathcal{F}^{I})_{N_{1}\dots N_{\ell}}(\mathcal{F}^{J})^{N_{1}\dots
N_{\ell}}g_{MN}%
\end{align}
We shall consider the following ansatz for the metric:
\begin{equation}
ds^{2}=\epsilon_{0}e^{2V}d\tau^{2}+e^{2U_{1}}ds^{2}(\mathcal{M}_{1}
)+e^{2U_{2}}ds^{2}(\mathcal{M}_{2})
\end{equation}
where $\mathcal{M}_{1}$, $\mathcal{M}_{2}$ are $n_{1}$ and $n_{2}$ dimensional
Einstein manifolds with co-ordinates $x^{\mu}$, $\mu=1,\dots,n_{1}$; $x^{a}$,
$a=1,\dots n_{2}$ respectively, also $d=1+n_{1}+n_{2}$ and $\epsilon_{0}=\pm
1$. We take
\begin{equation}
ds^{2}(\mathcal{M}_{1})=\eta_{\mu\nu}dx^{\mu}dx^{\nu},\qquad ds^{2}
(\mathcal{M}_{2})=h_{ab}dx^{a}dx^{b}%
\end{equation}
where $\eta_{\mu\nu}$ depends only on the $x^{\rho}$ co-ordinates, and
$h_{ab}$ depends only on the $x^{c}$ co-ordinates. The Ricci tensors of
$\mathcal{M}_{1}$, $\mathcal{M}_{2}$ are
\begin{equation}
(R^{1})_{\mu\nu}=k_{1}\eta_{\mu\nu},\qquad(R^{2})_{ab}=k_{2}h_{ab}%
\end{equation}
for constants $k_{1},k_{2}$. We also take $V=V(\tau)$, $U_{1}=U_{1}(\tau)$,
$U_{2}=U_{2}(\tau)$.

The metric signature is mostly positive. We set $\epsilon_{i}=1$,
$\epsilon_{i}=-1$ according as to whether $\mathcal{M}_{i}$ is Riemannian or
pseudo-Riemannian respectively, for $i=1,2$. So Lorentzian solutions have
exactly one of $\epsilon_{0},\epsilon_{1},\epsilon_{2}$ to be $-1$, with the
remaining two equal to $+1$. Instanton solutions have $\epsilon_{0}%
=\epsilon_{1}=\epsilon_{2}=1$.

We remark that the $\tau$ co-ordinate can be chosen in such a way that
\begin{equation}
V=c+n_{1}U_{1}+n_{2}U_{2} \label{coord1}%
\end{equation}
for constant $c$. With this choice, the non-zero components of the
$d$-dimensional Ricci tensor simplify somewhat, and one obtains
\begin{align}
R_{\tau\tau}  &  =-\partial_{\tau}^{2}V+\left(  \partial_{\tau}V\right)
^{2}-n_{1}\left(  \partial_{\tau}U_{1}\right)  ^{2}-n_{2}\left(
\partial_{\tau}U_{2}\right)  ^{2}\nonumber\\
R_{\mu\nu}  &  =\left(  k_{1}-\epsilon_{0}e^{2(U_{1}-V)}\partial_{\tau}
^{2}U_{1}\right)  \eta_{\mu\nu}\nonumber\\
R_{ab}  &  =\left(  k_{2}-\epsilon_{0}e^{2(U_{2}-V)}\partial_{\tau}^{2}
U_{2}\right)  h_{ab} \ .
\end{align}

\subsection{Electric Solutions}

For electric solutions, we take the $\mathcal{F}^{I}$ to be $(n_{1}+1)$-forms
with
\begin{equation}
\mathcal{F}^{I}=q^{I}(\tau)d\tau\wedge\mathrm{dvol}(\mathcal{M}_{1}).
\end{equation}
In particular, it will be useful to define
\begin{equation}
\mathcal{F}={\tilde{Q}}_{IJ}(\mathcal{F}^{I})_{N_{1}\dots N_{\ell}
}(\mathcal{F}^{J})^{N_{1}\dots N_{\ell}}=\epsilon_{0}\epsilon_{1}
(n_{1}+1)!e^{-2(V+n_{1}U_{1})}{\tilde{Q}}_{IJ}q^{I}q^{J}%
\end{equation}

Then the Einstein equations simplify to
\begin{equation}
-\partial_{\tau}^{2}V+\left(  \partial_{\tau}V\right)  ^{2}-n_{1}\left(
\partial_{\tau}U_{1}\right)  ^{2}-n_{2}\left(  \partial_{\tau}U_{2}\right)
^{2}=\mathcal{G}_{ij}{\partial_{\tau}\phi^{i}}{\partial_{\tau}\phi^{j}}
+{\frac{\epsilon_{0}(1-n_{2})}{2(n_{1}+1)!(1-n_{1}-n_{2})}}e^{2V}
\mathcal{F}\ , \label{ein1}%
\end{equation}
and
\begin{equation}
k_{1}-\epsilon_{0}e^{2(U_{1}-V)}\partial_{\tau}^{2}U_{1}={\frac{(1-n_{2}
)}{2(n_{1}+1)!(1-n_{1}-n_{2})}}e^{2U_{1}}\mathcal{F}\ , \label{ein2}%
\end{equation}
and
\begin{equation}
k_{2}-\epsilon_{0}e^{2(U_{2}-V)}\partial_{\tau}^{2}U_{2}={\frac{n_{1}}
{2(n_{1}+1)!(1-n_{1}-n_{2})}}e^{2U_{2}}\mathcal{F} \ . \label{ein3}%
\end{equation}
On eliminating $\mathcal{F}$ from ({\ref{ein2}}) and ({\ref{ein3}}) one
obtains
\begin{equation}
n_{1}k_{1}e^{2(V-U_{1})}+(n_{2}-1)k_{2}e^{2(V-U_{2})}-\epsilon_{0}
\partial_{\tau}^{2}(n_{1}U_{1}+(n_{2}-1)U_{2})=0. \label{meq1}%
\end{equation}
We shall further simplify this equation by setting
\begin{equation}
k_{1}=0
\end{equation}
so that on making the definition,
\begin{equation}
\sigma=n_{1}U_{1}+(n_{2}-1)U_{2} \label{seq1}%
\end{equation}
and using ({\ref{coord1}}), the condition ({\ref{meq1}}) is equivalent to
\begin{equation}
\partial_{\tau}^{2}\sigma=\epsilon_{0}(n_{2}-1)k_{2}e^{2c}e^{2\sigma}.
\label{meq1b}%
\end{equation}
This equation can be integrated up to obtain
\begin{equation}
\left(  \partial_{\tau}\sigma\right)  ^{2}=\epsilon_{0}(n_{2}-1)e^{2c}
(k_{2}e^{2\sigma}+\chi)
\end{equation}
for constant $\chi$. There are then two possibilities. In the first case, if
$k_{2}$ and $\chi$ are not both zero, then it is convenient to change
co-ordinates from $\tau$ to $\sigma$, and the metric can be written as
\begin{equation}
ds^{2}=e^{2U_{2}}\left(  {\frac{1}{(n_{2}-1)}}{\frac{e^{2\sigma}}
{k_{2}e^{2\sigma}+\chi}}d\sigma^{2}+ds^{2}(\mathcal{M}_{2})\right)
+e^{2U_{1}}ds^{2}(\mathcal{M}_{1}) \label{mmet1}%
\end{equation}
where $U_{1}=U_{1}(\sigma)$ and $U_{2}=U_{2}(\sigma)$ are related by
({\ref{seq1}}).

The remaining content of the Einstein equations can be rewritten, eliminating
$V$ and $U_{2}$ in favour of $U_{1}$:
\begin{equation}
k_{2}e^{2\sigma}\partial_{\sigma}U_{1}+(k_{2}e^{2\sigma}+\chi)\partial
_{\sigma}^{2}U_{1}={\frac{1}{2(n_{1}+1)!(1-n_{1}-n_{2})}}e^{{\frac
{2n_{2}\sigma}{n_{2}-1}}-{\frac{2n_{1}U_{1}}{n_{2}-1}}}\mathcal{F}
\label{ex1a}%
\end{equation}
and
\begin{align}
n_{2}\chi-n_{1}(n_{1}+n_{2}-1)(k_{2}e^{2\sigma}+\chi)\left(  \partial_{\sigma
}U_{1}\right)  ^{2}  &  =(n_{2}-1)(k_{2}e^{2\sigma}+\chi)\mathcal{G}
_{ij}\partial_{\sigma}{\phi}^{i}\partial_{\sigma}{\phi}^{j}\nonumber\\
&  +{\frac{1}{2(n_{1}+1)!}}e^{{\frac{2n_{2}\sigma}{n_{2}-1}}-{\frac
{2n_{1}U_{1}}{n_{2}-1}}}\mathcal{F} \ . \label{ex11b}%
\end{align}

In the second case, if $k_{2}=\chi=0$, then the metric is
\begin{equation}
ds^{2}=e^{-\frac{2n_{1}U}{n_{2}-1}}\left(  \epsilon_{0}d\tau^{2}
+ds^{2}(\mathcal{M}_{2})\right)  +e^{2U}ds^{2}(\mathcal{M}_{1}) \label{mmet2}%
\end{equation}
where $U(\tau)$ satisfies
\begin{equation}
\partial_{\tau}^{2}U=-{\frac{\epsilon_{0}(1-n_{2})}{2(n_{1}+1)!(1-n_{1}
-n_{2})}}e^{\frac{2(c+n_{1}U_{1})}{1-n_{2}}}\mathcal{F} \label{ex2a}%
\end{equation}
and
\begin{equation}
{\frac{n_{1}(n_{1}+n_{2}-1)}{1-n_{2}}}\left(  \partial_{\tau}U\right)
^{2}=\mathcal{G}_{ij}{\partial_{\tau}\phi^{i}}{\partial_{\tau}\phi^{j}}
+{\frac{\epsilon_{0}}{2(n_{1}+1)!}}e^{\frac{2(c+n_{1}U_{1})}{1-n_{2}}
}\mathcal{F}\ .
\end{equation}

\subsection{Magnetic Solutions}

For magnetic solutions, we take the $\mathcal{F}^{I}$ to be $n_{2}$-forms
with
\begin{equation}
\mathcal{F}^{I}=p^{I}\mathrm{dvol}(\mathcal{M}_{2})
\end{equation}
for constant $p^{I}$. Again it is useful to define
\begin{equation}
\mathcal{F}={\tilde{Q}}_{IJ}(\mathcal{F}^{I})_{N_{1}\dots N_{\ell}
}(\mathcal{F}^{J})^{N_{1}\dots N_{\ell}}=\epsilon_{2}(n_{2})!e^{-2n_{2}U_{2}
}{\tilde{Q}}_{IJ}p^{I}p^{J}\ .
\end{equation}
The Einstein equations are then
\begin{equation}
-\partial_{\tau}^{2}V+\left(  \partial_{\tau}V\right)  ^{2}-n_{1}\left(
\partial_{\tau}U_{1}\right)  ^{2}-n_{2}\left(  \partial_{\tau}U_{2}\right)
^{2}={\mathcal{G}}_{ij}{\partial_{\tau}\phi^{i}}{\partial_{\tau}\phi^{j}
}+{\frac{\epsilon_{0}(n_{2}-1)}{2(n_{2})!(1-n_{1}-n_{2})}}e^{2V}\mathcal{F}%
\end{equation}
and
\begin{equation}
k_{1}-\epsilon_{0}e^{2(U_{1}-V)}\partial_{\tau}^{2}U_{1}={\frac{(n_{2}
-1)}{2(n_{2})!(1-n_{1}-n_{2})}}e^{2U_{1}}\mathcal{F} \label{mein5}%
\end{equation}
and
\begin{equation}
k_{2}-\epsilon_{0}e^{2(U_{2}-V)}\partial_{\tau}^{2}U_{2}=-{\frac{n_{1}
}{2(n_{2})!(1-n_{1}-n_{2})}}e^{2U_{2}}\mathcal{F}\ . \label{mein6}%
\end{equation}
Again, on eliminating $\mathcal{F}$ from between ({\ref{mein5}}) and
({\ref{mein6}}), one obtains the condition ({\ref{meq1}}). So, on setting
$k_{1}=0$ and defining $\sigma$ as in ({\ref{seq1}}), the metric can again be
written as ({\ref{mmet1}}) or ({\ref{mmet2}}). The Einstein equations become
\begin{equation}
k_{2}e^{2\sigma}\partial_{\sigma}U_{1}+(k_{2}e^{2\sigma}+\chi)\partial
_{\sigma}^{2}U_{1}=-{\frac{1}{2(n_{2})!(1-n_{1}-n_{2})}}e^{\frac{2}{n_{2}
-1}\left(  {n_{2}\sigma}-{n_{1}U_{1}}\right)  }\mathcal{F}%
\end{equation}
and
\begin{align}
&  n_{2}\chi-(k_{2}e^{2\sigma}+\chi)\left(  n_{1}(n_{1}+n_{2}-1)\left(
\partial_{\sigma}U_{1}\right)  ^{2}+(n_{2}-1)\mathcal{G}_{ij}\partial_{\sigma
}\phi^{i}\partial_{\sigma}\phi^{j}\right) \nonumber\\
&  =-{\frac{1}{2(n_{2})!}}e^{\frac{2}{n_{2}-1}\left(  {n_{2}\sigma}
-{n_{1}U_{1}}\right)  }\mathcal{F} \ .
\end{align}

In the special case $k_{1}=k_{2}=\chi=0$, \ then the magnetically charged
solution is given by%

\begin{equation}
ds^{2}=e^{-2\frac{n_{1}U}{n_{2}-1}}\left(  \epsilon_{0}d\tau^{2}
+ds^{2}(\mathcal{M}_{2})\right)  +e^{2U}ds^{2}(\mathcal{M}_{1})
\end{equation}
where in these cases both $\mathcal{M}_{1}$and $\mathcal{M}_{2}$ are
Ricci-flat manifolds. The Einstein's equations reduce to
\begin{align}
-\frac{n_{1}\left(  n_{1}+n_{2}-1\right)  \left(  \partial_{\tau}U\right)
^{2}}{\left(  n_{2}-1\right)  }  &  ={\mathcal{G}}_{ij}{\partial_{\tau}
\phi^{i}}{\partial_{\tau}\phi^{j}}-{\frac{\epsilon_{0}}{2(n_{2})!}}
e^{-2\frac{n_{1}U}{n_{2}-1}}\mathcal{F}\nonumber\\
\partial_{\tau}^{2}U  &  =-{\frac{(n_{2}-1)\epsilon_{0}}{2(n_{2}
)!(1-n_{1}-n_{2})}}e^{-2\frac{n_{1}U}{n_{2}-1}}\mathcal{F} \ .
\end{align}

\section{Four-Dimensional Examples}

In this section we will consider four-dimensional examples. For convenience,
we shall first recall some aspects of special geometry \cite{speone} for
Lorentzian and Euclidean signatures. The Euclidean versions of the special
geometries can be obtained from the standard counterparts by replacing $i$ by
$e$ which satisfies $e^{2}=1$ and $\bar{e}=-e$. All fields in the Euclidean
theory can be written in the form
\begin{equation}
\phi=\operatorname{Re}\phi+e\operatorname{Im}\phi,\text{ \ }\bar{\phi
}=\operatorname{Re}\phi-e\operatorname{Im}\phi.
\end{equation}
Further details on para-complex geometry, para-holomorphic bundles,
para-K\"{a}hler manifolds and affine special para-K\"{a}hler manifolds can be
found in \cite{mohaupt1}. The theory of ungauged $N=2$, $D=4$ supergravity
theories can be described by the Lagrangian
\begin{equation}
\mathcal{L}_{4}=\sqrt{|g|}\left[  R-2Q_{IJ}\partial_{\mu}X^{I}\partial^{\mu
}{\bar{X}}^{J}+\frac{1}{2}\left(  \operatorname{Im}\mathcal{N}_{IJ}
\mathcal{F}^{I}\cdot\mathcal{F}^{J}+\operatorname{Re}\mathcal{N}
_{IJ}\mathcal{F}^{I}\cdot\mathcal{\tilde{F}}^{J}\right)  \;\right]  .
\label{Action}%
\end{equation}
The theory has $n+1$ gauge fields $A^{I},(\mathcal{F}^{I}=d\mathcal{A}^{I})$
and $n$ scalars $z^{a}.$ The scalar fields are complex for $(1,3)$ signature
and para-complex for $(0,4)$ signature
\begin{equation}
z^{a}=\operatorname{Re}z^{a}+i_{\epsilon}\operatorname{Im}z^{a}%
\end{equation}
Here $i_{\epsilon}=i$ for the $(1,3)$ signature and $i_{\epsilon}=e$ for
$(0,4)$ signature, $i_{\epsilon}^{2}=\epsilon$. We also have
\begin{equation}
Q_{IJ}\partial_{\mu}X^{I}\partial^{\mu}{\bar{X}}^{J}=g_{a\bar{b}}\partial
_{\mu}z^{a}\partial^{\mu}\bar{z}^{b}%
\end{equation}
where$\ g_{a\bar{b}}=\partial_{a}\partial_{\bar{b}}K$ is the K\"{a}hler metric
and $K$ is the K\"{a}hler potential.

The coordinates $X^{I}$ are related to the covariantly holomorphic sections
\begin{align}
V  &  =\left(
\begin{array}
[c]{c}%
L^{I}\\
M_{I}%
\end{array}
\right)  ,\text{ \ \ \ }I=0,...,n\nonumber\\
\mathcal{D}_{\bar{a}}V  &  =\left(  \partial_{\bar{a}}-\frac{1}{2}
\partial_{\bar{a}}K\right)  V=0.
\end{align}
obeying the constraint
\begin{equation}
i_{\epsilon}\langle V,\bar{V}\rangle=i_{\epsilon}\left(  \bar{L}^{I}
M_{I}-L^{I}\bar{M}_{I}\right)  =1,
\end{equation}
by
\begin{equation}
\Omega=e^{-K/2}V=\left(
\begin{array}
[c]{c}%
X^{I}\\
F_{I}%
\end{array}
\right)  ,\text{ \ \ \ \ }\partial_{\bar{a}}\Omega=0.
\end{equation}
The K\"{a}hler potential is given by
\begin{equation}
e^{-K}=i_{\epsilon}\left(  \bar{X}^{I}F_{I}-X^{I}\bar{F}_{I}\right)  .
\end{equation}
We also have the relations
\begin{align}
M_{I}  &  =\mathcal{N}_{IJ}L^{J},\text{ \ \ \ \ \ }\mathcal{D}_{a}
M_{I}=\mathcal{\bar{N}}_{IJ}\mathcal{D}_{a}L^{I}\nonumber\\
\mathcal{D}_{a}  &  =\left(  \partial_{a}+\frac{1}{2}\partial_{a}K\right)  .
\end{align}
Note that the $N=2$ supergravity models can be described in terms of a
holomorphic homogeneous prepotential $F=F(X^{I})$ of degree two. In this case
$F_{I}=\frac{\partial F}{\partial X^{I}}$, $F_{IJ}=\frac{\partial F}{\partial
X^{I}\partial X^{J}}$, etc. We note the following useful relations%

\begin{align}
F  &  ={\frac{1}{2}}F_{I}X^{I},\text{ \ \ \ }F_{I}=F_{IJ}X^{J},\nonumber\\
\text{ \ \ }X^{I}F_{IJK}  &  =0,\nonumber\\
X^{I}F_{IJKL}  &  =-F_{JKL},\nonumber\\
F_{I}\partial_{\mu}X^{I}-X^{I}\partial_{\mu}F_{I}  &  =0,\nonumber\\
g^{a\bar{b}}\mathcal{D}_{a}L^{I}\mathcal{D}_{\bar{b}}\bar{L}^{J}  &
=-\frac{1}{2}\left(  \text{Im}\mathcal{N}\right)  ^{IJ}-\bar{L}^{I}L^{J}\ .
\label{useful}%
\end{align}
The scalar and gauge couplings are given by \cite{mohaupt3}
\begin{align}
\mathcal{N}_{IJ}  &  =\bar{F}_{IJ}-\epsilon{\frac{i_{\epsilon}}{(XNX)}
}(NX)_{I}(NX)_{J}\nonumber\\
Q_{IJ}  &  =e^{K}N_{IJ}+e^{2K}(N{\bar{X}})_{I}(NX)_{J}%
\end{align}
where
\begin{equation}
N_{IJ}=i_{\epsilon}\left(  \bar{F}_{IJ}-F_{IJ}\right)  ,
\end{equation}
and we use the notation $(NX)_{I}=N_{IJ}X^{J}$, $(N{\bar{X}})_{I}=N_{IJ}%
{\bar{X}}^{J}$, $XNX=X^{I}X^{J}N_{IJ}$.

\subsection{Four-Dimensional Electric Solutions}

The electrically charged solutions are characterised by real scalar fields
$X^{I}$ and purely imaginary prepotential $F$. Using the general results of
the previous section, taking $n_{1}=1$ and $n_{2}=2$; when $k_{2}$ and $\chi$
are not both zero, the four-dimensional metric can be written in the form%

\begin{equation}
ds^{2}=e^{-2U}\left[  {\frac{e^{4\sigma}}{k_{2}e^{2\sigma}+\chi}}d\sigma
^{2}+e^{2\sigma}ds^{2}(\mathcal{M}_{2})\right]  +\epsilon_{1}e^{2U}d\rho^{2}%
\end{equation}
where we have set $U=U_{1},U_{2}=\sigma-U$, and $\mathcal{M}_{2}$ is a
two-dimensional Einstein manifold. The gauge fields are $F^{I}=F_{\sigma\rho
}^{I}d\sigma\wedge d\rho$, with gauge field equations
\begin{equation}
\partial_{\sigma}\left(  e^{-2U}\sqrt{|k_{2}e^{2\sigma}+\chi|}
\operatorname{Im}\mathcal{N}_{IJ}F_{\sigma\rho}^{I}\right)  =0\ .
\end{equation}
It follows that
\begin{align}
\operatorname{Im}\mathcal{N}_{IJ}\mathcal{F}_{\sigma\rho}^{J}  &
=\frac{e^{2U}}{\sqrt{\left\vert k_{2}e^{2\sigma}+\chi\right\vert }}
q_{I}\nonumber\\
\operatorname{Im}\mathcal{N}_{IJ}\mathcal{F}^{I\sigma\rho}\mathcal{F}
_{\sigma\rho}^{J}  &  =\epsilon_{1}\epsilon_{0}\operatorname{Im}
\mathcal{N}^{IJ}e^{4U-4\sigma}q_{I}q_{J}%
\end{align}
for constant $q_{I}$. Moreover, the Einstein equations reduce to%

\begin{equation}
X^{I}\bigg((k_{2}e^{2\sigma}+\chi)\partial_{\sigma}^{2}\operatorname{Im}
F_{I}+k_{2}e^{2\sigma}\partial_{\sigma}\operatorname{Im}F_{I}-\chi
\operatorname{Im}F_{I}\bigg)=0 \label{fi4}%
\end{equation}
and
\begin{equation}
{\operatorname{Im}}\mathcal{N}^{IJ}\bigg[(k_{2}e^{2\sigma}+\chi)\partial
_{\sigma}\operatorname{Im}F_{I}\partial_{\sigma}\operatorname{Im}F_{J}
-\chi\operatorname{Im}F_{I}\operatorname{Im}F_{J}-\frac{\epsilon_{1}
\epsilon_{0}}{4\epsilon}q_{I}q_{J}\bigg]=0 \label{sec4}%
\end{equation}
where we have used the relations coming from special geometry
\begin{align}
Q_{IJ}\partial_{\sigma}X^{I}\partial_{\sigma}X^{J}  &  =\frac{1}{2F}\left(
\frac{\left(  \partial_{\sigma}F\right)  ^{2}}{2F}-\partial_{\sigma}
^{2}F+X^{I}\partial_{\sigma}^{2}F_{I}\right) \nonumber\\
\mathcal{N}_{IJ}\partial_{\sigma}X^{I}\partial_{\sigma}X^{J}  &
=\mathcal{N}^{IJ}\partial_{\sigma}F_{I}\partial_{\sigma}F_{J}={\frac{\left(
\partial_{\sigma}F\right)  ^{2}}{F}}-\partial_{\sigma}^{2}F+X^{I}
\partial_{\sigma}^{2}F_{I} \label{nicere}%
\end{align}
as well as the ansatz
\begin{equation}
e^{-2U}=4i_{\epsilon}F\ . \label{ansatz}%
\end{equation}
The equations (\ref{fi4}) and (\ref{sec4}) can be solved by taking%

\begin{equation}
2\operatorname{Im}F_{I}=A_{I}\sqrt{\vert k_{2}+\chi e^{-2\sigma} \vert}
+B_{I}e^{-\sigma}%
\end{equation}
for constants $A_{I}, B_{J}$, which must satisfy the relation%

\begin{equation}
{\operatorname{Im}}\mathcal{N}^{IJ}\left(  k_{2}\left(  \epsilon_{0}\chi
A_{I}A_{J}-B_{I}B_{J}\right)  +\frac{\epsilon_{1}\epsilon_{0}}{\epsilon}
q_{I}q_{J}\right)  =0 \ .
\end{equation}

Turning to the scalar equations of motion, this gives for our ansatz, after
some calculation, the following equation
\begin{equation}
\partial_{a}\mathrm{{Im}}{\mathcal{N}}^{IJ}\left(  k_{2}\left(  \epsilon
_{0}\chi A_{I}A_{J}-B_{I}B_{J}\right)  +\frac{\epsilon_{0}\epsilon_{1}%
}{\epsilon}q_{I}q_{J}\right)  =0\ .
\end{equation}
As special cases, one can consider the solutions for which $k_{2}\neq0$ and
$\chi=0,$%

\begin{equation}
ds^{2}=e^{-2U}\left[  {\frac{1}{k_{2}}}d\sigma^{2}+e^{2\sigma}ds^{2}
(\mathcal{M}_{2})\right]  +\epsilon_{1}e^{2U}d\rho^{2}%
\end{equation}
then we find
\begin{equation}
\operatorname{Im}F_{I} =A_{I}+B_{I}e^{-\sigma}%
\end{equation}
for constants $A_{I}, B_{I}$, and the remaining content of the Einstein and
scalar equations is
\begin{equation}
{\operatorname{Im}}\mathcal{N}^{IJ}\left(  B_{I}B_{J}-\frac{\epsilon
_{1}\epsilon_{0}}{4k_{2}\epsilon}q_{I}q_{J}\right)  =0 \ ,
\end{equation}
and
\begin{equation}
\partial_{a} {\operatorname{Im}}\mathcal{N}^{IJ}\left(  B_{I}B_{J}%
-\frac{\epsilon_{1}\epsilon_{0}}{4k_{2}\epsilon}q_{I}q_{J}\right)  =0 \ .
\end{equation}

Another special class of solutions for $k_{2}=0$ and $\chi\neq0,$
\begin{equation}
ds^{2}=e^{-2U}\left[  {\frac{e^{4\sigma}}{\chi}}d\sigma^{2}+e^{2\sigma}
ds^{2}(\mathcal{M}_{2})\right]  +\epsilon_{1}e^{2U}d\rho^{2}\ .
\end{equation}
then the solution is given by%

\begin{equation}
\operatorname{Im}F_{I} =A_{I}e^{\sigma}+B_{I}e^{-\sigma}%
\end{equation}
for constants $A_{I}, B_{I}$, and the remaining content of the Einstein and
scalar equations is%

\begin{equation}
{\operatorname{Im}}\mathcal{N}^{IJ}\left(  2\left(  B_{J}A_{I}+B_{I}
A_{J}\right)  +\frac{\epsilon_{1}\epsilon_{0}}{4\chi\epsilon}q_{I}
q_{J}\right)  =0 \ ,
\end{equation}
and
\begin{equation}
\partial_{a} {\operatorname{Im}}\mathcal{N}^{IJ} \left(  2\left(  B_{J}%
A_{I}+B_{I} A_{J}\right)  +\frac{\epsilon_{1}\epsilon_{0}}{4\chi\epsilon}q_{I}
q_{J}\right)  =0 \ .
\end{equation}

The remaining class of electrically charged solution, with $k_{1}=k_{2}%
=\chi=0$, is given by
\begin{equation}
ds^{2}=e^{-2U}\left(  \epsilon_{0}d\tau^{2}+ds^{2}(\mathcal{M}_{2})\right)
+\epsilon_{1}e^{2U}d\rho^{2}\ .
\end{equation}
In this case, the remaining content of the Einstein equations, again written
in terms of $U$, is
\begin{align}
\partial_{\tau}^{2}U  &  ={\frac{\epsilon_{1}}{2}}e^{2U}\operatorname{Im}
\mathcal{N}^{IJ}q_{I}q_{J}\nonumber\\
-\left(  \partial_{\tau}U\right)  ^{2}+\partial_{\tau}^{2}U  &  =Q_{IJ}
{\partial_{\tau}X^{I}}{\partial_{\tau}X^{J}} \ .
\end{align}
Again using the relations (\ref{nicere}) but with the derivatives taken with
respect to $\tau$, together with the ansatz (\ref{ansatz}), the Einstein
equations are equivalent to%

\begin{align}
\operatorname{Im}\mathcal{N}^{IJ}\left(  \partial_{\tau}\operatorname{Im}
F_{I}\partial_{\tau}\operatorname{Im}F_{J}-{\frac{\epsilon\epsilon_{1}}{4}
}q_{I}q_{J}\right)   &  =0\nonumber\\
X^{I}\partial_{\tau}^{2}F_{I}  &  =0 \ .
\end{align}
We solve these equations by taking
\begin{equation}
\mathrm{Im}F_{I}=A_{I}+B_{I}\tau
\end{equation}
for constants $A_{I},B_{I}$ so the Einstein equations reduce to
\begin{equation}
\mathrm{Im}{\mathcal{N}}^{IJ}\left(  B_{I}B_{J}-{\frac{1}{4}}\epsilon
\epsilon_{1}q_{I}q_{J}\right)  =0 \ .
\end{equation}
The scalar equation also reduces to%

\begin{equation}
\partial_{a}\operatorname{Im}\mathcal{N}^{IJ}\left(  \partial_{\tau
}\operatorname{Im}F_{I}\partial_{\tau}\operatorname{Im} F_{J}-\frac{1}%
{4}\epsilon\epsilon_{1}q_{I}q_{J}\right)  =0.\
\end{equation}
which gives the condition
\begin{equation}
\partial_{a}\operatorname{Im}\mathcal{N}^{IJ}\left(  B_{I}B_{J}-{\frac{1}{4}%
}\epsilon\epsilon_{1}q_{I}q_{J}\right)  =0 \ .
\end{equation}



\subsection{Four-Dimensional Magnetic Solutions}

For magnetic solutions with $k_{2}$ and $\chi$ not both zero, we take the both
the scalars $X^{I}$ and the prepotential $F$ to be purely imaginary, with
$n_{1}=1$ and $n_{2}=2$. We then find from the Einstein equations%

\begin{equation}
F_{I}\left[  k_{2}e^{2\sigma}\partial_{\sigma}\operatorname{Im}X^{I}%
+(k_{2}e^{2\sigma}+\chi)\partial_{\sigma}^{2}\operatorname{Im}X^{I}%
-\chi\operatorname{Im}X^{I}\right]  =0
\end{equation}
and
\begin{equation}
\operatorname{Im}\mathcal{N}_{IJ}\left[  \chi\operatorname{Im}X^{J}%
\operatorname{Im}X^{I}-(k_{2}e^{2\sigma}+\chi)\partial_{\sigma}%
\operatorname{Im}X^{I}\partial_{\sigma}\operatorname{Im}X^{J}+\frac
{\epsilon_{2}}{4}p^{I}p^{J}\ \right]  =0
\end{equation}
where we have taken
\begin{equation}
e^{-2U}=-4i_{\epsilon}F
\end{equation}
The first equation can be solved by taking
\begin{equation}
2\operatorname{Im}X^{I}=A^{I}\sqrt{|k_{2}+\chi e^{-2\sigma}|}+B^{I}e^{-\sigma}%
\end{equation}

The remaining content of the Einstein and scalar equations can then be written
in terms of the constants $A^{I},B^{I}$ as%

\begin{equation}
\mathrm{Im}{\mathcal{N}}_{IJ}\left(  k_{2}(\epsilon_{0}\chi A^{I}A^{J}%
-B^{I}B^{J})+\epsilon_{2}p^{I}p^{J}\right)  =0\ ,
\end{equation}
and
\begin{equation}
\partial_{a}\mathrm{Im}{\mathcal{N}}_{IJ}\left(  k_{2}(\epsilon_{0}\chi
A^{I}A^{J}-B^{I}B^{J})+\epsilon_{2}p^{I}p^{J}\right)  =0.
\end{equation}

A special class of solutions is when $k_{2}\neq0$ and $\chi=0,$ then a
solution is given by

\begin{align}
\operatorname{Im}X^{I}  &  =A^{I}+B^{I}e^{-\sigma}\nonumber\\
\operatorname{Im}\mathcal{N}_{IJ}\left(  B^{I}B^{J}-\frac{1}{4k_{2}}%
\epsilon_{2}p^{I}p^{J}\right)   &  =0\nonumber\\
\partial_{a}\operatorname{Im}\mathcal{N}_{IJ}\left(  B^{I}B^{J}-\frac
{1}{4k_{2}}\epsilon_{2}p^{I}p^{J}\right)   &  =0\ .
\end{align}

Another special class of solutions is when $k_{2}=0$ and $\chi\neq0$, for
which a solution is given by%

\begin{align}
\operatorname{Im}X^{I}  &  =A^{I}e^{\sigma}+B^{I}e^{-\sigma}\nonumber\\
\operatorname{Im}\mathcal{N}_{IJ}\left( A^{I} B^{J} + A^{J} B^{I}
+{\frac{\epsilon_{2} }{8 \chi}} p^{I} p^{J} \right)   &  =0\nonumber\\
\partial_{a}\operatorname{Im}\mathcal{N}_{IJ}\left(  A^{I} B^{J} + A^{J} B^{I}
+{\frac{\epsilon_{2} }{8 \chi}} p^{I} p^{J} \right)   &  =0\ .
\end{align}

For the class of magnetic solutions where $k_{1}=k_{2}=\chi=0,$ then the
solution is given by
\begin{equation}
ds^{2}=\epsilon_{0}e^{-2U}d\tau^{2}+e^{2U}d\rho^{2}+e^{-2U}ds^{2}%
(\mathcal{M}_{2})
\end{equation}
Taking the ansatz
\begin{equation}
e^{-2U}=-4i_{\epsilon}F\
\end{equation}
the Einstein equations imply
\begin{align}
F_{I}\partial_{\tau}^{2}\operatorname{Im}X^{I}  &  =0\nonumber\\
\operatorname{Im}\mathcal{N}_{IJ}\left(  \partial_{\tau}\operatorname{Im}%
X^{I}\partial_{\tau}\operatorname{Im}X^{J}-{\frac{\epsilon_{0}\epsilon_{2}}%
{4}}p^{I}p^{J}\right)   &  =0\ .
\end{align}
The first of these equations can be satisfied by setting
\begin{equation}
\operatorname{Im}X^{I}=A^{I}+B^{I}\tau
\end{equation}

for real constants $A^{I},B^{I}$. The remaining content of the Einstein
equations, and of the scalar equations, is then given by%

\begin{equation}
\mathrm{Im}{\mathcal{N}}_{IJ}\left(  B^{I}B^{J}-{\frac{1}{4}}\epsilon
_{0}\epsilon_{2}p^{I}p^{J}\right)  =0\ ,
\end{equation}
and
\begin{equation}
\partial_{a}\mathrm{Im}{\mathcal{N}}_{IJ}\left(  B^{I}B^{J}-{\frac{1}{4}%
}\epsilon_{0}\epsilon_{2}p^{I}p^{J}\right)  =0.
\end{equation}


\section{Five-Dimensional Examples}

The Lagrangian of the five-dimensional $N=2$ supergravity theory with an
arbitrary number of vector multiplets is given by \cite{GST}%

\begin{equation}
\mathcal{L}_{5}=\sqrt{|g|}\left[  R-\frac{1}{2}Q_{IJ}F_{\mu\nu}^{I}F^{\mu\nu
J}-g_{ij}\partial_{\mu}\phi^{i}\partial^{\mu}\phi^{j}-\frac{1}{24}
\epsilon^{\mu\nu\rho\sigma\tau}C_{IJK}F_{\mu\nu}^{I}F_{\rho\sigma}^{J}A_{\tau
}^{K}\right]  .
\end{equation}
In very special geometry, one has the fields $X^{I}=X^{I}(\phi),$ the so
called special coordinates satisfying
\begin{equation}
\mathcal{V=}\text{ }X^{I}X_{I}={\frac{1}{6}}C_{IJK}X^{I}X^{J}X^{K}=1
\label{cn}%
\end{equation}
where, $X_{I}$ are the dual coordinates. The gauge coupling metric $Q_{IJ}$
and the metric $g_{ij}$ depend on the scalar fields via the relations
\begin{align}
Q_{IJ}  &  ={\frac{9}{2}}X_{I}X_{J}-{\frac{1}{2}}C_{IJK}X^{K}\nonumber\\
g_{ij}  &  =Q_{IJ}\frac{\partial X^{I}}{\partial\phi^{i}}\frac{\partial X^{J}
}{\partial\phi^{j}}|_{\mathcal{V}=1}.
\end{align}
We also note the useful relations
\begin{equation}
Q_{IJ}X^{J}={\frac{3}{2}}X_{I}\,,\qquad Q_{IJ}\partial_{i}X^{J}=-{\frac{3}{2}
}\partial_{i}X_{I}\,.
\end{equation}
The gauge and scalar equations of motion are
\begin{equation}
\partial_{\mu}\left(  \sqrt{|g|}Q_{IJ}F^{I\mu\nu}\right)  +\frac{1}
{16}\epsilon^{\nu\rho\sigma\mu\tau}C_{IJK}F_{\rho\sigma}^{J}F_{\mu\tau} ^{K}=0
\label{max}%
\end{equation}
and
\begin{equation}
\sqrt{|g|}\partial_{i}Q_{IJ}\left[  \frac{1}{2}F_{\mu\nu}^{I}F^{\mu\nu
J}+\partial_{\mu}X^{I}\partial^{\mu}X^{J}\right]  -2\partial_{\mu}\left(
\sqrt{|g|}Q_{IJ}\partial^{\mu}X^{J}\right)  \partial_{i}X^{I}=0\ .
\label{scal}%
\end{equation}

\subsection{Five-Dimensional Electric solutions}

For electrically charged solutions, with $n_{1}=1,n_{2}=3$, the metric can be
written as
\begin{equation}
ds^{2}=e^{\sigma-U}\left[  {\frac{e^{2\sigma}}{2\left(  k_{2}e^{2\sigma}
+\chi\right)  }}d\sigma^{2}+ds^{2}(\mathcal{M}_{2})\right]  +\epsilon
_{1}e^{2U}d\rho^{2}\ .
\end{equation}
The gauge fields are given by $F^{I}=F_{\sigma\rho}^{I}d\sigma\wedge d\rho$,
and the gauge equations imply
\begin{equation}
\partial_{\sigma}\left(  e^{-2U}\sqrt{\left\vert k_{2}e^{2\sigma}
+\chi\right\vert }Q_{IJ}F_{\sigma\rho}^{I}\right)  =0\ .
\end{equation}
We thus have
\begin{equation}
e^{-2U}\sqrt{\left\vert k_{2}e^{2\sigma}+\chi\right\vert }Q_{IJ}F_{\sigma\rho
}^{I}=-\frac{1}{2}q_{I}%
\end{equation}
for constants $q_{I}$. The Einstein equations are given by%

\begin{align}
k_{2}e^{2\sigma}\partial_{\sigma}U+(k_{2}e^{2\sigma}+\chi)\partial_{\sigma
}^{2}U  &  =-\frac{\epsilon_{0}\epsilon_{1}}{6}e^{2U}{Q}^{IJ}q_{I}
q_{J}\nonumber\\
\chi-(k_{2}e^{2\sigma}+\chi)\left(  \partial_{\sigma}U\right)  ^{2}  &
=\frac{2}{3}(k_{2}e^{2\sigma}+\chi)Q_{IJ}\partial_{\sigma}X^{I}\partial
_{\sigma}X^{J}+\frac{\epsilon_{0}\epsilon_{1}}{6}e^{2U}{Q}^{IJ}q_{I}q_{J} \ .
\end{align}
We shall set
\begin{equation}
X_{I}=\frac{1}{3}e^{U}f_{I}.
\end{equation}
Then from the equations of very special geometry (\ref{cn}) we obtain
\begin{equation}
e^{-U}=\frac{1}{3}f_{I}X^{I}.
\end{equation}
Moreover using very special geometry we obtain the following relations%

\begin{align}
Q_{IJ}\partial_{\sigma}X^{I}\partial_{\sigma}X^{J}  &  =\frac{3}{2}
X^{I}\partial_{\sigma}^{2}X_{I}=\frac{1}{2}\left(  3\partial_{\sigma}
^{2}U-3\left(  \partial_{\sigma}U\right)  ^{2}+X^{I}e^{U}\partial_{\sigma}
^{2}f_{I}\right) \nonumber\label{ali}\\
Q_{IJ}\partial_{\sigma}X^{I}\partial_{\sigma}X^{J}  &  =\frac{9}{4}
Q^{IJ}\partial_{\sigma}X_{I}\partial_{\sigma}X_{J}=\frac{1}{2}\left(
-3\left(  \partial_{\sigma}U\right)  ^{2}+\frac{1}{2}Q^{IJ}e^{2U}
\partial_{\sigma}f_{I}\partial_{\sigma}f_{J}\right)  .
\end{align}
Then the Einstein equations can be rewritten as%

\begin{align}
X^{I}\left(  k_{2}e^{2\sigma}\partial_{\sigma}f_{I}+(k_{2}e^{2\sigma}
+\chi)\partial_{\sigma}^{2}f_{I}-\chi f_{I}\right)   &  =0\nonumber\\
Q^{IJ}\left(  (k_{2}e^{2\sigma}+\chi)\partial_{\sigma}f_{I}\partial_{\sigma
}f_{J}-\chi f_{I}f_{J}+\epsilon_{0}\epsilon_{1}q_{I}q_{J}\right)   &  =0
\end{align}
which can be solved by taking%

\begin{equation}
f_{I}=A_{I}\sqrt{\vert k_{2}+\chi e^{-2\sigma}\vert}+B_{I}e^{-\sigma}%
\end{equation}
for constants $A_{I}, B_{I}$, with the condition%

\begin{equation}
Q^{IJ}\left[  k_{2}\left(  \chi\epsilon_{0} A_{I}A_{J}-B_{I}B_{J}\right)
-\epsilon_{0}\epsilon_{1}q_{I}q_{J}\right]  =0 \ .
\end{equation}
Using the relations of very special geometry, and in particular
\begin{equation}
\partial_{i}X^{I}=\frac{3}{4}\partial_{i}Q^{IJ}X_{J}%
\end{equation}
then the scalar equation of motion gives%

\begin{equation}
\partial_{i}Q^{IJ}\left[  -{\frac{\epsilon_{0}\epsilon_{1}}{k_{2}e^{2\sigma
}+\chi}}q_{I}q_{J}-\partial_{\sigma}f_{I}\partial_{\sigma}f_{J}+\left(
\partial_{\sigma}^{2}f_{I}\right)  f_{J}+\frac{k_{2}e^{2\sigma}}{\left(
k_{2}e^{2\sigma}+\chi\right)  }\partial_{\sigma}f_{I}f_{J}\right]  =0 \ .
\end{equation}
This reduces to%

\begin{equation}
\partial_{i}Q^{IJ}\left[  k_{2}\left(  \epsilon_{0}\chi A_{I}A_{J}-B_{I}
B_{J}\right)  -\epsilon_{0}\epsilon_{1}q_{I}q_{J}\right]  =0 \ .
\end{equation}
For the special case when $k_{2}=0,\chi\neq0,$ we obtain the solution
\begin{align}
f_{I}  &  =A_{I}e^{\sigma}+B_{I}e^{-\sigma}\nonumber\\
Q^{IJ}\left(  A_{I}B_{J}+A_{J}B_{I}-\frac{\epsilon_{0}\epsilon_{1}}{2\chi
}q_{I}q_{J}\right)   &  =0\nonumber\\
\partial_{i}Q^{IJ}\left(  A_{I}B_{J}+A_{J}B_{I}-\frac{\epsilon_{0}\epsilon
_{1}}{2\chi}q_{I}q_{J}\right)   &  =0 \ .
\end{align}
For the case when $k_{2}\neq0,\chi=0,$ we obtain the solution%

\begin{align}
f_{I}  &  =A_{I}+B_{I}e^{-\sigma}\nonumber\\
Q^{IJ}\left(  B_{I}B_{J}+\frac{\epsilon_{0}\epsilon_{1}}{k_{2}}q_{I}
q_{J}\right)   &  =0\nonumber\\
\partial_{i}Q^{IJ}\left(  B_{I}B_{J}+\frac{\epsilon_{0}\epsilon_{1}}{k_{2}
}q_{I}q_{J}\right)   &  =0 \ .
\end{align}
For the cases of electric solutions with $k_{1}=k_{2}=\chi=0$, the metric is
given by
\begin{equation}
ds^{2}=e^{-U}\left(  \epsilon_{0}d\tau^{2}+ds^{2}(\mathcal{M}_{2})\right)
+e^{2U}ds^{2}(\mathcal{M}_{1}) \ .
\end{equation}
In this case, the content of the Einstein equations is given by
\begin{align}
-{\frac{3}{2}}\left(  \partial_{\tau}U\right)  ^{2}+\frac{3}{2}\partial_{\tau
}^{2}U  &  =Q_{IJ}{\partial_{\tau}X}^{I}{\partial_{\tau}X}^{J}\nonumber\\
\partial_{\tau}^{2}U  &  =-{\frac{\epsilon_{0}}{6}}e^{-U}\mathcal{F} \ .
\end{align}
On setting
\begin{equation}
X_{I}=\frac{1}{3}e^{U}f_{I}\nonumber
\end{equation}
\begin{equation}
Q_{IJ}F_{\sigma\rho}^{I}=-\frac{e^{2U}}{2}q_{I} \ ,
\end{equation}
and using special geometry, we have the following conditions
\begin{align}
X^{I}\partial_{\tau}^{2}f_{I}  &  =0\nonumber\\
Q^{IJ}\left(  \partial_{\sigma}f_{I}\partial_{\sigma}f_{J}+\epsilon_{1}
q_{I}q_{J}\right)   &  =0 \ .
\end{align}
These equations are solved by
\begin{equation}
f_{I}=A_{I}+B_{I}\tau
\end{equation}
and the remaining content of the Einstein and scalar equations is
\begin{align}
Q^{IJ}\left(  B_{I}B_{J}+\epsilon_{1}q_{I}q_{J}\right)   &  =0\nonumber\\
\partial_{i} Q^{IJ}\left(  B_{I}B_{J}+\epsilon_{1}q_{I}q_{J}\right)   &  =0
\ .
\end{align}

\subsection{Five-Dimensional Magnetic solutions}

\bigskip For magnetic solutions, taking $n_{1}=n_{2}=2$, the $F^{I}$ are
$2$-forms with
\begin{equation}
F^{I}=p^{I}\mathrm{dvol}(\mathcal{M}_{2})
\end{equation}
for constant $p^{I}$. In this case, the metric is
\begin{equation}
ds^{2}=e^{-4U}\left[  {\frac{e^{4\sigma}}{k_{2}e^{2\sigma}+\chi}}d\sigma
^{2}+e^{2\sigma}ds^{2}(\mathcal{M}_{2})\right]  +e^{2U}ds^{2}(\mathcal{M}
_{1}) \ .
\end{equation}
The Einstein equations are then given by
\begin{align}
k_{2}e^{2\sigma}\partial_{\sigma}U+(k_{2}e^{2\sigma}+\chi)\partial_{\sigma
}^{2}U  &  ={\frac{1}{3}}e^{{4U}}\epsilon_{2}{Q}_{IJ}p^{I}p^{J}\nonumber\\
2\chi-6(k_{2}e^{2\sigma}+\chi)\left(  \partial_{\sigma}U\right)  ^{2}  &
=(k_{2}e^{2\sigma}+\chi)Q_{IJ}\partial_{\sigma}X^{I}\partial_{\sigma}
X^{J}-\epsilon_{2}e^{4U}{Q}_{IJ}p^{I}p^{J} \ .
\end{align}
We also set
\begin{equation}
X^{I}=e^{2U}f^{I}. \label{an}%
\end{equation}
Using very special geometry we obtain from the Einstein equations%

\begin{equation}
X_{I}\left(  (k_{2}e^{2\sigma}+\chi)\partial_{\sigma}^{2}f^{I}+k_{2}
e^{2\sigma}\partial_{\sigma}f^{I}-\frac{4}{3}\chi f^{I}\right)  =0
\label{magone}%
\end{equation}
and
\begin{equation}
Q_{IJ}\left(  (k_{2}e^{2\sigma}+\chi)\partial_{\sigma}f^{I}\partial_{\sigma
}f^{J}-\frac{4}{3}\chi f^{I}f^{J}-\epsilon_{2}p^{I}p^{J}\right)  =0 \ .
\label{magtwo}%
\end{equation}

Furthermore, the scalar field equations, after making use of several very
special geometry relations, can be written as
\begin{equation}
\partial_{i}Q_{IJ}\left(  \epsilon_{2}p^{I}p^{J}-(k_{2}e^{2\sigma}
+\chi)\partial_{\sigma}f^{I}\partial_{\sigma}f^{J}+k_{2}e^{2\sigma}
f^{I}\partial_{\sigma}f^{J}+(k_{2}e^{2\sigma}+\chi)f^{I}\partial_{\sigma}
^{2}f^{J}\right)  =0 \ . \label{magsc1}%
\end{equation}
We solve ({\ref{magone}}) by setting
\begin{equation}
(k_{2}e^{2\sigma}+\chi)\partial_{\sigma}^{2}f^{I}+k_{2}e^{2\sigma}
\partial_{\sigma}f^{I}-\frac{4}{3}\chi f^{I}=0 \label{odemag1}%
\end{equation}
and on substituting ({\ref{odemag1}}) into ({\ref{magsc1}}) the remaining
conditions are ({\ref{magtwo}}) together with ({\ref{magsc1}}), which can be
rewritten as
\begin{equation}
\partial_{i}Q_{IJ}\left(  (k_{2}e^{2\sigma}+\chi)\partial_{\sigma}
f^{I}\partial_{\sigma}f^{J}-\frac{4}{3}\chi f^{I}f^{J}-\epsilon_{2}p^{I}
p^{J}\right)  =0 \ . \label{magtwob}%
\end{equation}

The solution to ({\ref{odemag1}}) depends on the sign of $\epsilon_{0} \chi$.
If $\epsilon_{0} \chi>0$ then we find%

\begin{equation}
f^{I}=A^{I}\left(  J\right)  ^{-\sqrt{\frac{1}{3}}}+B^{I}\left(  J\right)
^{+\sqrt{\frac{1}{3}}} \label{magsolv1}%
\end{equation}
for real constants $A^{I},B^{I}$, where
\begin{equation}
J={\frac{\left(  1+\sqrt{1+{\frac{k_{2}}{\chi}}e^{2\sigma}}\right)  }{\left(
1-\sqrt{1+{\frac{k_{2}}{\chi}}e^{2\sigma}}\right)  }} \ .
\end{equation}
On substituting ({\ref{magsolv1}}) into ({\ref{magtwo}}), and ({\ref{magtwob}
}) we find%

\begin{equation}
Q_{IJ}\left[  \frac{8}{3}\chi\left(  A^{I}B^{J}+B^{I}A^{J}\right)
+\epsilon_{2}p^{I}p^{J}\right]  =0
\end{equation}
and
\begin{equation}
\partial_{i}Q_{IJ}\left[  \frac{8}{3}\chi\left(  A^{I}B^{J}+B^{I}A^{J}\right)
+\epsilon_{2}p^{I}p^{J}\right]  =0 \ .
\end{equation}

If however $\epsilon_{0}\chi<0$ then one finds
\begin{equation}
f^{I}=A^{I}K^{\frac{1}{\sqrt{3}}}+{\bar{A}}^{I}K^{-{\frac{1}{\sqrt{3}}}}
\label{negmagsolv1}%
\end{equation}
for complex constants $A^{I}$, where
\begin{equation}
K={\frac{\left[  \sqrt{-1-{\frac{k_{2}}{\chi}}e^{2\sigma}}-i\right]  }{\left[
\sqrt{-1-{\frac{k_{2}}{\chi}}e^{2\sigma}}+i\right]  }} \ .
\end{equation}
On substituting ({\ref{negmagsolv1}}) into ({\ref{magtwo}}), and
({\ref{magtwob}}) we find
\begin{equation}
Q_{IJ}\left[  \frac{8}{3}\chi\left(  A^{I}{\bar{A}}^{J}+{\bar{A}}^{I}
A^{J}\right)  +\epsilon_{2}p^{I}p^{J}\right]  =0
\end{equation}
and
\begin{equation}
\partial_{i}Q_{IJ}\left[  \frac{8}{3}\chi\left(  A^{I}{\bar{A}}^{J}+{\bar{A}
}^{I}A^{J}\right)  +\epsilon_{2}p^{I}p^{J}\right]  =0 \ .
\end{equation}

For $\chi=0,k_{2}\neq0,$ the solution is given by%

\begin{align}
f^{I}  &  =A^{I}+B^{I}e^{-\sigma}\nonumber\\
Q_{IJ}\left(  B^{I}B^{J}-\frac{\epsilon_{2}}{k_{2}}p^{I}p^{J}\right)   &
=0\nonumber\\
\partial_{i}Q_{IJ}\left(  B^{I}B^{J}-\frac{\epsilon_{2}}{k_{2}}p^{I}
p^{J}\right)   &  =0 \ ,
\end{align}
for real constants $A^{I}, B^{I}$.

For $\chi\neq0,k_{2}=0,$ the solution is given by%

\begin{align}
f^{I}  &  =A^{I}e^{-\sqrt{\frac{4}{3}}\sigma}+B^{I}e^{\sqrt{\frac{4}{3}}
\sigma}\nonumber\\
Q_{IJ}\left[  \frac{8}{3}\left(  A^{I}B^{J}+B^{J}A^{I}\right)  +\frac
{\epsilon_{2}}{\chi}p^{I}p^{J}\right]   &  =0\nonumber\\
\partial_{i}Q_{IJ}\left[  \frac{8}{3}\left(  A^{I}B^{J}+B^{J}A^{I}\right)
+\frac{\epsilon_{2}}{\chi}p^{I}p^{J}\right]   &  =0 \ ,
\end{align}
for real constants $A^{I}, B^{I}$.

Finally the magnetic solutions with both $\mathcal{M}_{1}\mathcal{\ }$and
$\mathcal{M}_{2}$ Ricci-flat and $\chi=0,$ are given by%

\begin{equation}
ds^{2}=e^{-4U}\left(  \epsilon_{0}d\tau^{2}+ds^{2}(\mathcal{M}_{2})\right)
+e^{2U}ds^{2}(\mathcal{M}_{1}).
\end{equation}
In this case Einstein's equations reduce to
\begin{align}
-6\left(  \partial_{\tau}U\right)  ^{2}+3\partial_{\tau}^{2}U  &
=Q_{IJ}\partial_{\tau}X^{I}\partial_{\tau}X^{J}\nonumber\\
\partial_{\tau}^{2}U  &  ={\frac{\epsilon_{0}}{3}}e^{4U}\epsilon_{2}{Q}
_{IJ}p^{I}p^{J} \ .
\end{align}
Again using the ansatz
\begin{equation}
X^{I}=e^{2U}f^{I},
\end{equation}
these equations reduce to%

\begin{align}
X_{I}^{2}\partial_{\tau}^{2}f^{I}  &  =0\nonumber\\
Q_{IJ}\left(  \partial_{\tau}f^{I}\partial_{\tau}f^{J}-{\epsilon_{0}}
\epsilon_{2}p^{I}p^{J}\right)   &  =0
\end{align}
which can be solved by
\begin{equation}
f^{I}=A^{I}+B^{I}\tau
\end{equation}
and the remaining content of the Einstein and scalar equations is
\begin{equation}
Q_{IJ}\left(  B^{I}B^{J}-{\epsilon_{0}}\epsilon_{2}p^{I}p^{J}\right)  =0 \ ,
\end{equation}
and
\begin{equation}
\partial_{i} Q_{IJ}\left(  B^{I}B^{J}-{\epsilon_{0}}\epsilon_{2}p^{I}%
p^{J}\right)  =0 \ .
\end{equation}

\bigskip

\vskip 0.3cm

\noindent\textbf{Acknowledgements} \vskip0.1cm JG is supported by the STFC
grant, ST/1004874/1. JG would like to thank the Department of Mathematical
Sciences, University of Liverpool for hospitality during which part of this
work was completed. The work of W.S. is supported in part by the National
Science Foundation under grant number PHY-1415659.


\vskip 0.5cm

\begin{thebibliography}{99}                                                                                               %


\bibitem {recentlower}D. Klemm and E. Zorzan, \textit{All null supersymmetric
backgrounds of $N=2$, $D=4$ gauged supergravity coupled to abelian vector
multiplets}, Class. Quant. Grav. \textbf{26} (2009) 145018; arXiv:0902.4186
[hep-th]. S. L. Cacciatori, D. Klemm, D. S. Mansi and E. Zorzan, \textit{All
timelike supersymmetric solutions of $N=2$, $D=4$ gauged supergravity coupled
to abelian vector multiplets}, JHEP \textbf{05} (2008) 097; arXiv:0804.0009
[hep-th]. J. Grover, J. B. Gutowski, C. A. R. Herdeiro, P. Meessen, A.
Palomo-Lozano and W. A. Sabra, \textit{Gauduchon-Tod structures, Sim holonomy
and De Sitter supergravity}, JHEP \textbf{07} (2009) 069; arXiv:0905.3047
[hep-th]. J. B. Gutowski and W. A. Sabra, \textit{Solutions of Minimal Four
Dimensional de Sitter Supergravity}, Class. Quantum Grav. 27 (2010) 235017;
arXiv:0903.0179 [hep-th]. J. Grover, J. B. Gutowski, C. A. R. Herdeiro and W.
A. Sabra, \textit{HKT Geometry and de Sitter Supergravity, } Nucl. Phys.
\textbf{B809} (2009) 406; arXiv:0806.2626 [hep-th]. J. Grover, J. B. Gutowski
and W. A. Sabra, \textit{Null Half-Supersymmetric Solutions in
Five-Dimensional Supergravity,} JHEP \textbf{10} (2008) 103; arXiv:0802.0231
[hep-th]. U. Gran, J. Gutowski and G. Papadopoulos, \textit{Geometry of all
supersymmetric four-dimensional N = 1 supergravity backgrounds}, JHEP
\textbf{06} (2008); arXiv:0802.1779 [hep-th].

\bibitem {Tod}K. P. Tod, \textit{All Metrics Admitting Supercovariantly
Constant Spinors}, Phys. Lett. \textbf{B121} (1983) 241.

\bibitem {IWP}Z. Perj\'{e}s, \textit{Solutions of the coupled Einstein-Maxwell
equations representing the fields of spinning sources}, Phys. Rev. Lett.
\textbf{27} (1971) 1668; W. Israel and G. A. Wilson, \textit{A Class of
stationary electromagnetic vacuum fields}, J. Math. Phys. \textbf{13} (1972) 865.

\bibitem {mp}S. D. Majumdar, \textit{A Class of Exact Solutions of Einstein's
Field Equations}, Phys. Rev. \textbf{72} (1947) 930; A. Papapetrou, Proc. Roy.
Irish. Acad. A51 (1947) 191.

\bibitem {sab}S. Ferrara, R. Kallosh and A. Strominger, \textit{N=2 Extremal
Black Holes}, Phys. Rev. \textbf{D52} (1995) 5412; hep-th/9508072. S. Ferrara
and R. Kallosh, \textit{Supersymmetry and Attractors}, Phys. Rev. \textbf{D54}
(1996) 1514; hep-th/9602136. W. A. Sabra, \textit{General static N = 2 black
holes}, Mod. Phys. Lett. \textbf{A12} (1997) 2585; hep-th/9703101. W. A.
Sabra,\textit{\ Black holes in N = 2 supergravity and harmonic functions},
Nucl. Phys. \textbf{B510} (1998) 247; hep-th/9704147.

\bibitem {BLS}K. Behrndt, D. Lust and W. A. Sabra, \textit{Stationary
solutions of N=2 supergravity}, Nucl. Phys. \textbf{B510} (1998) 264; hep-th/9705169.

\bibitem {ortin}P. Meessen and T. Ort\'{\i}n, \textit{The supersymmetric
configurations of $N=2$, $d=4$ supergravity coupled to vector supermultiplets}
, Nucl. Phys. \textbf{B749} (2006) 291; hep-th/0603099.

\bibitem {sabrafive}W. A. Sabra, \textit{General BPS Black Holes In Five
Dimensions}, Mod. Phys. Lett. \textbf{A13} (1998); hep-th/9708103.

\bibitem {sabrachams5}A. H. Chamseddine and W. A. Sabra, \textit{Metrics
Admitting Killing Spinors in Five-Dimensions}, Phys. Lett. \textbf{B426}
(1998) 36; hep-th/9801161.

\bibitem {mag}A. H. Chamseddine and W. A. Sabra, \textit{Calabi-Yau Black
Holes and Enhancement of Supersymmetry in Five Dimensions}, Phys. Lett.
\textbf{B460} (1999) 63; hep-th/9903046.

\bibitem {unique5}H. S. Reall, {\textit{Higher dimensional black holes and
supersymmetry,}} Phys.\ Rev.\ \textbf{D68} (2003) 024024
[Erratum-ibid.\ \textbf{D70} (2004) 089902]; hep-th/0211290. Jan B. Gutowski,
\textit{Uniqueness of Five-Dimensional Supersymmetric Black Holes}, JHEP
\textbf{08} (2004) 049; hep-th/0404079.

\bibitem {5dsys}J. P. Gauntlett, J. B. Gutowski, C. M. Hull, S. Pakis and H.
S. Reall, \textit{All supersymmetric solutions of minimal supergravity in five
dimensions}, Class. Quant. Grav. \textbf{20} (2003) 4587; hep-th/0209114.

\bibitem {BCS}K. Behrndt, M. Cvetic and W. A. Sabra, \textit{Non-Extreme Black
Holes of Five-Dimensional N=2 AdS Supergravity}, Nucl. Phys. \textbf{B553
}(1999) 317; hep-th/9810227.

\bibitem {nonextremaljan}J. B. Gutowski, W. A. Sabra, \textit{Five Dimensional
Non-Supersymmetric Black Holes and Strings}, JHEP \textbf{05} (2009) 092;
arXiv:0803.3189 [hep-th].

\bibitem {GO4}P. Galli, T. Ort\'{\i}n, J. Perz and C. S. Shahbazi,
\textit{Non-Extremal Black Holes of $N=2$, $d=4$ Supergravity}, JHEP
\textbf{07} (2011) 041; arXiv:1105.3311 [hep-th]. P. Meesen and T.~Ort\'{\i}n,
\textit{Non-Extremal Black Holes of N=2, d=5 Supergravity}, Phys.Lett.
\textbf{{B707}} (2012) 178; arXiv:1107.5454 [hep-th].

\bibitem {DH07}M. Dunajski and S. A. Hartnoll, \textit{Einstein-Maxwell
gravitational instantons and five dimensional solitonic strings}, Class.
Quantum. Grav. \textbf{24}, (2007) 1841; hep-th/0610261.

\bibitem {instantons}J. B. Gutowski and W. A. Sabra, \textit{Gravitational
Instantons and Euclidean Supersymmetry,} Phys. Lett. \textbf{{B693}} (2010)
498; arXiv:1007.2421 [hep-th].

\bibitem {instanton}J. B. Gutowski and W. A. Sabra, \textit{Para-Complex
Geometry and Gravitational Instantons,} Class. Quant. Grav. \textbf{{30}}
(2013) 195001; arXiv:1210.2332 [hep-th].

\bibitem {mohaupt3}V. Cortes and T. Mohaupt, \textit{Special Geometry of
Euclidean Supersymmetry III: the local r-map, instantons and black holes},
JHEP \textbf{07} (2009) 066; arXiv:0905.2844 [hep-th].

\bibitem {eusy}J. B. Gutowski, W. A. Sabra, \textit{\ Euclidean N=2
Supergravity,} Phys. Lett. \textbf{B718} (2012) 610; arXiv:1209.2029 [hep-th].

\bibitem {speone}S. Cecotti, B. Craps, F. Roose, W. Troost and A. Van Proeyen,
\textit{What is Special K\"{a}hler Geometry}?, Nucl. Phys. \textbf{B503}
(1997); hep-th/9703082.

\bibitem {mohaupt1}V. Cortes, C. Mayer and T. Mohaupt and F. Saueressig,
\textit{Special Geometry of Euclidean Supersymmetry I: Vector Multiplets,}
arXiv:hep-th/031200, JHEP \textbf{03} (2004) 028; hep-th/0312001.

\bibitem {GST}M. Gunaydin, G. Sierra and P. K. Townsend, \textit{The Geometry
of N = 2 Maxwell Einstein Supergravity and Jordan Algebras,} Nucl. Phys.
\textbf{B242} (1984) 244. B. de Wit and A. van Proyen, \textit{Broken sigma
model isometries in very special geometry}, Phys. Lett. \textbf{B293} (1992)
94; hep-th/9207091. A. C. Cadavid, A. Ceresole, R. D'Auria, and S. Ferrara,
\textit{Eleven dimensional supergravity compactified on Calabi-Yau
threefolds}, \textit{Phys. Lett.} \textbf{B357} (1995) 76; hep-th/9506144.
\end{thebibliography}
\end{document}